\documentclass[nofootinbib,apsspec,twocolumn,groupedaddress,lengthcheck,preprintnumbers]{revtex4}
\usepackage{amsmath}
\usepackage{amssymb}
\usepackage{graphics}
\usepackage[dvips]{graphicx}
\usepackage{graphicx}
\usepackage[usenames]{color}

\newcommand \beq{\begin{eqnarray}}
\newcommand \eeq{\end{eqnarray}}
\def\simge{\mathrel{%
       \rlap{\raise 0.511ex \hbox{$>$}}{\lower 0.511ex \hbox{$\sim$}}}}
\def\simle{\mathrel{
       \rlap{\raise 0.511ex \hbox{$<$}}{\lower 0.511ex \hbox{$\sim$}}}}

\usepackage[colorlinks=true,linktocpage=true,linkcolor=blue,citecolor=blue]{hyperref}

\begin{document}
\title{Testing dark decays of baryons in neutron stars}
\author{Gordon Baym,$^a$ D.\ H.\ Beck,$^a$  Peter Geltenbort,$^b$ and Jessie Shelton$^a$} 
\affiliation{\mbox{$^a$Department of Physics, University of Illinois, 1110
  W. Green Street, Urbana, IL 61801} \\
\mbox{$^b$Institut Max von Laue Paul Langevin, 71 avenue des Martyrs, F-38042 Grenoble Cedex 9, France}\\
}

\begin{abstract}

 The observation of neutron stars with masses
  greater than one solar mass places severe demands on any exotic
  neutron decay mode that could explain the discrepancy between beam
  and bottle measurements of the neutron lifetime.  If the neutron can
  decay to a stable, feebly-interacting dark fermion, the maximum
  possible mass of a neutron star is $0.7 M_\odot$, while all
  well-measured neutron star masses exceed one $M_\odot$.  The
existence of $2M_\odot$ neutron stars further indicates that any
  explanation beyond the Standard Model for the neutron lifetime
  puzzle requires dark matter to be part of a multi-particle dark
  sector with highly constrained interactions.  Beyond the
  neutron lifetime puzzle, our results indicate that neutron stars
  provide unique and useful probes of GeV-scale dark sectors
  coupled to the Standard Model via baryon-number-violating interactions.

\end{abstract}

\maketitle

The neutron lifetime anomaly, the discrepancy in the beam \cite{Yue:2013qrc, Byrne:1996zz}
vs.~bottle \cite{Serebrov:2017bzo, Pattie:2017vsj,Arzumanov:2015tea,
  Steyerl:2012zz, Pichlmaier:2010zz, Serebrov:2004zf, Mampe:1993an}
measurements of the lifetime of the neutron, is a long-standing puzzle
\cite{greene, sciam}.  Briefly, the bottle technique, an inclusive
measurement of the neutron lifetime, yields $\tau_{bottle} = 879.6 \pm 0.6$
s, which is discrepant at the $4\sigma$ level with the exclusive
measurement of the neutron lifetime via beam experiments,
$\tau_{beam} = 888.0 \pm 2.0$ s \cite{PDG}.
In a recent paper Fornal and Grinstein \cite{grinstein} made the
intriguing suggestion that new decay channels of the neutron, $n$, in particular
\beq 
n \to \chi + \gamma, \quad n\to \chi + e^+e^-, \quad n \to \chi +\phi,
\label{decays}
\eeq 
where $\chi$ is a dark matter fermion, $\phi$ is a dark matter boson, and $\gamma$
is a photon, could explain the shorter lifetime in the bottle
experiments.  
The amplitude for these processes must be
sufficiently large to allow a rate, $\Gamma \sim10^{-5}$ s$^{-1}$,
to explain the bottle-beam anomaly.\footnote{This resolution to the neutron decay puzzle faces a number of challenges.  The $n\to \chi +\gamma$
decay mode has been tested in a recent experiment \cite{tang}
that excluded all
branching ratios that could account for the lifetime anomaly.   In
addition, Ref.~\cite{cms} argues that recent measurements of the axial
renormalization constant, $g_A$, likely point to a shorter Standard
Model (SM) lifetime of the neutron, more in line with the bottle results.} 

     With this suggestion in mind we show that neutron stars are powerful laboratories to test proposed dark decays of baryons.
The conversion of baryons to dark fermions through processes of the form in Eq.~(\ref{decays}) lead, in the absence of strong self-interactions of the dark fermions, $\chi$, to a maximum neutron star mass much smaller than observed masses.  Thus the existence of neutron stars with masses up to 2$M_\odot$ \cite{OzelFreire,Demorest,Antoniadis2013} allows us to draw broad and generic conclusions about the type of baryon number-violating dark interactions of the neutron required for a Beyond the Standard Model  (BSM) explanation of the neutron lifetime puzzle.

\begin{figure}[h]
\includegraphics[width = 0.45\textwidth]{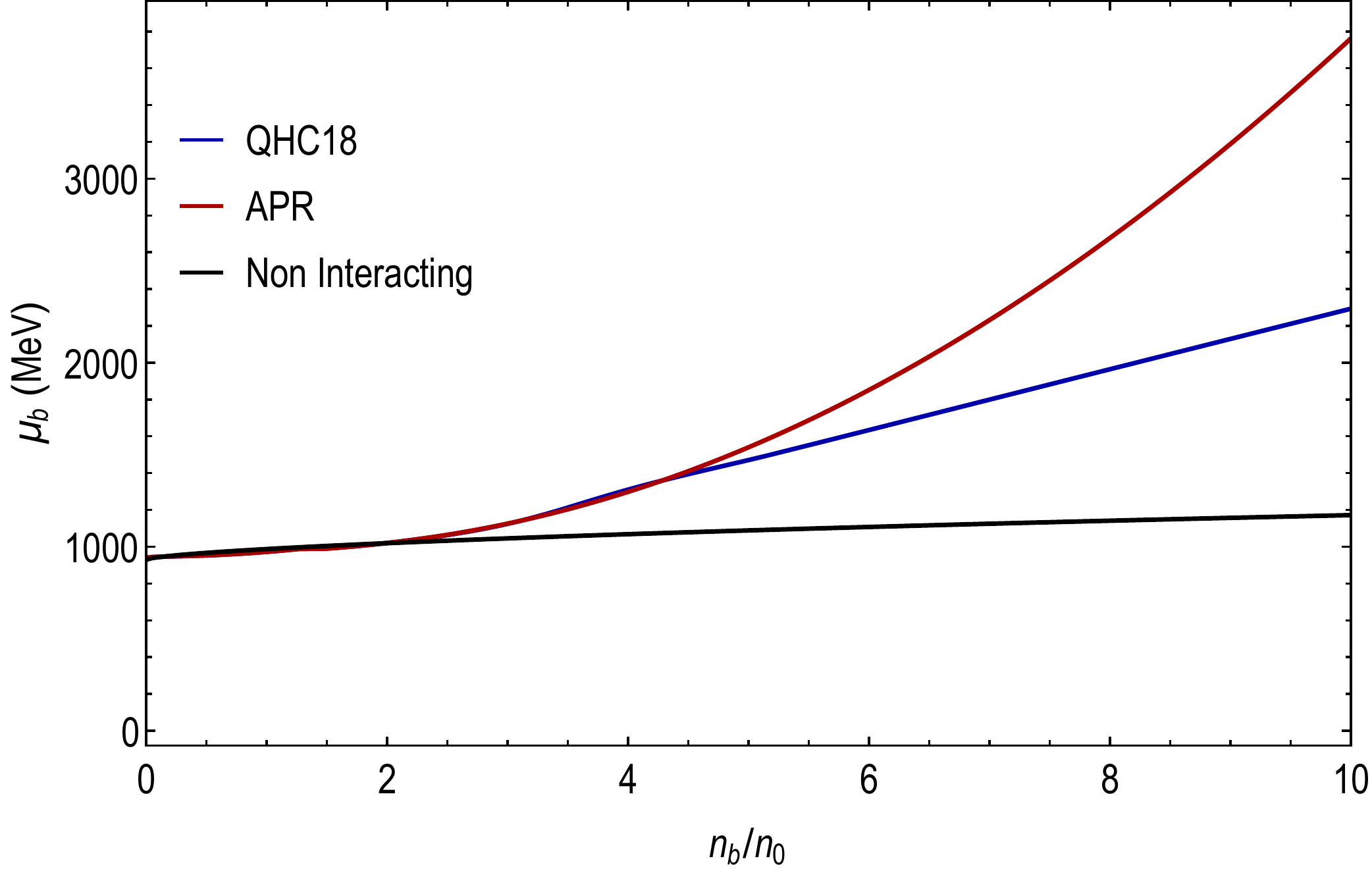}
\caption{\footnotesize Baryon (or neutron) chemical potential (including
  rest mass) in MeV vs. the baryon density in units of nuclear matter
  density, $n_0=0.16$ fm$^{-3}$, for the quark-hadron crossover
  equation of state, QHC18(0.8,1.5) \cite{nstar}, for the stiffer APR equation of state \cite{APR} and for non-interacting (free)
 neutrons.
This figure shows how much more expensive it would be for a baryon to
  remain at high densities instead of turning into a weakly
  interacting dark matter particle with $m_\chi < \mu_b$.}
\label{mub-nb}
\end{figure}

   The processes of Eq.~(\ref{decays}) would convert a fraction of the neutrons present into $\chi$s during the formation of a neutron star.
The $\chi$s would sit in the gravitational potential well of the neutron star, in thermodynamic equilibrium with the normal neutron star matter,\footnote{  The $10^5$ s time scale for $n\to \chi + Y$ needed to resolve the neutron lifetime puzzle is very short compared to inferred neutron star ages, which range from hundreds to billions of years \cite{ATNF}.}
and form a non-interacting Fermi gas, similar to a non-interacting neutron gas.  
The basic physics is that, except near nuclear matter density, the
interactions of neutrons with the neutron star medium are effectively
repulsive, and thus the conversion of a neutron into a weakly interacting dark matter
particle is generally highly energetically favored.
Figure~\ref{mub-nb} shows the baryon chemical potential, $\mu_b$, vs. the baryon density, $n_b$, in units of $n_0$, the nuclear matter saturation density, $\simeq$ 0.16 fm$^{-3}$, 
for the modern quark-hadron crossover [QHC18(0.8,1.5)] neutron star matter equation of state  \cite{nstar}, for the 'stiffer' Akmal-Pandharipande-Ravenhall (APR) equation of state \cite{APR},  and for free neutrons.\footnote{The QHC18 equations of state take quark degrees of freedom in the interior into account consistently and allow 2$M_\odot$ neutron stars.   They are in striking agreement with the equation of state constraints deduced by LIGO from the recent binary neutron star merger \cite{BNS}.  The ingredients of these equations of state are
effectively: i) the APR equation of state for nuclear matter in beta equilibrium, up to baryon
density $\simeq 2n_0$;  ii) above baryon density $\sim 5 n_0$ a quark
matter equation of state with a repulsive contact interaction between
the quarks with coupling constant $g_v$, equal here to $0.8 G$, and an
effective BCS pairing interaction between quarks with coupling
constant $H$, equal here to $1.5G$, where $G$ is the
Nambu--Jona-Lasinio quark-quark coupling constant; and iii) between
the two extremes a smooth interpolation of $P$ vs. $\mu_b$.}  For given $n_b$, the conversion of neutrons to free fermions of equal mass would generally gain of order hundreds of MeV per neutron. 
  
    The interactions (\ref{decays}) are phrased in terms of the neutron instead of the quarks comprising the neutron.   Thus to describe
the effects of these interactions on neutron stars it is simplest to use the language of neutron degrees of freedom, although the calculations we present are valid for more general baryon and quark degrees of freedom.  We calculate neutron star models in the presence of a generic interaction $n \to \chi + Y$,  where $Y$ is a possibly multi-particle final state with zero net chemical potential, $\mu_Y=0$.  Such interactions include the highly pertinent SM final states $Y = \gamma, e^+ e^-$ as well as a broad range of BSM possibilities such as a dark photon.  We assume for simplicity that the $\chi$ have spin 1/2.

   In a neutron star with non-interacting $\chi$s, a sea of neutrons, with Fermi momentum $k_b$ and density $n_b \equiv k_b^3/3\pi^2$,
would be in equilibrium with a sea of $\chi$'s with Fermi momentum $k_\chi$, density $n_\chi
= k_\chi^3/3\pi^2$,  and chemical potential $\mu_\chi^2 =\sqrt{m_\chi^2+k_\chi^2}$.  In equilibrium the $\chi$'s must have the same chemical potential, $\mu_b$ as that of the baryons.  To calculate the relative population of baryons and $\chi$ we write
$k_\chi= yk_b$ where $y$ is itself a function of $k_b$.  Although we use relativistic kinematics in the numerical results presented below, we provide here the non-relativistic limits to illustrate the physics most simply.  Non-relativistically, chemical equilibrium leads to
\beq
    y^2= \frac{m_\chi\left(\mu_b-m_\chi\right)}{m_n\left(\mu_n^0-m_n\right)}, \quad \mu_b \ge m_\chi.     
\eeq
where $\mu_n^0$ is the chemical potential of a free
neutron gas at density $k_b^3/3\pi^2$.  We take for simplicity  $m_\chi=m_n$ in addressing the Fornal and Grinstein proposition.   Given that $\mu_b > m_n$ at high densities we also show results for a range of $m_\chi>m_n$.

\begin{figure}[h]
\includegraphics[width = 0.45\textwidth]{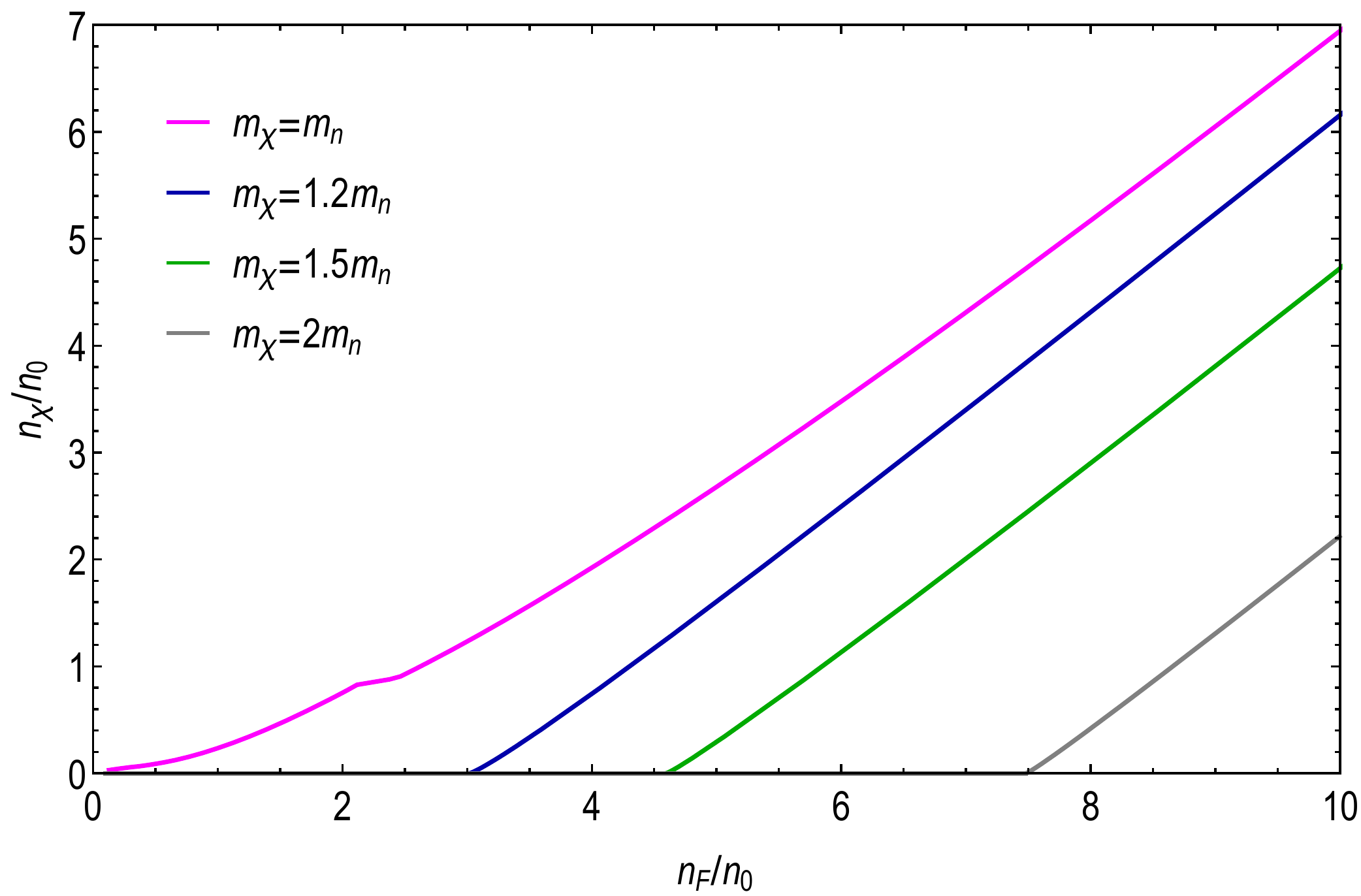}
\caption{\footnotesize{ Number density of dark matter fermions $\chi$ in chemical equilibrium
    as a function of the total number of fermions, $n_F=n_b+n_\chi$,
    in units of nuclear matter density, $n_0$, for the QHC18 neutron star equation of state and for different dark matter fermion masses: $m_\chi = m_n$, $m_\chi = 1.2m_n$, $m_\chi = 1.5m_n$, and $m_\chi = 2m_n$.   (The small flattening in the $m_\chi = m_n$ curve
reflects the onset of pion condensation in the APR equation of state \cite{APR}.)  }}
\label{x}
\end{figure}

   The total density of fermions is
$n_F=n_b+n_\chi = n_b(1+y^3)$.
For $m_\chi=m_n$, we find that at total fermion density $n_F \simeq n_0$ (nuclear matter saturation density), about 40\% of
the fermions are $\chi$s, while at $n_F \simeq 4n_0$ the number of
$\chi$ and normal baryons are approximately equal, and at $n_F=10n_0$,
$\sim$ 70\% of the fermions are $\chi$s (see Fig.~\ref{x}).  If the
baryon chemical potential is below $m_\chi$, no $\chi$
can be present and $y \equiv 0$.

\vspace{0.5cm}

The $\chi$s contribute only their rest mass and kinetic energy to
the total energy density, $\epsilon$, of the matter; again non-relativistically for illustration
\beq
    \epsilon = \epsilon_b(n_b) + m_\chi n_b y^3 + \frac{k_b^5}{10\pi^2 m_\chi}y^5,
\eeq
while the total pressure is
\beq
  P = P_b(n_b) +  \frac{k_b^5}{15\pi^2 m_\chi}y^5,
\eeq
where $\epsilon_b$ is the energy density and $P_b$ the pressure of normal matter.
The $\chi$s increase the energy density more than the
pressure, and thus at high densities soften the equation of state, and lower the maximum neutron star mass.
   
   \begin{figure}[h]
\includegraphics[width = 0.45\textwidth]{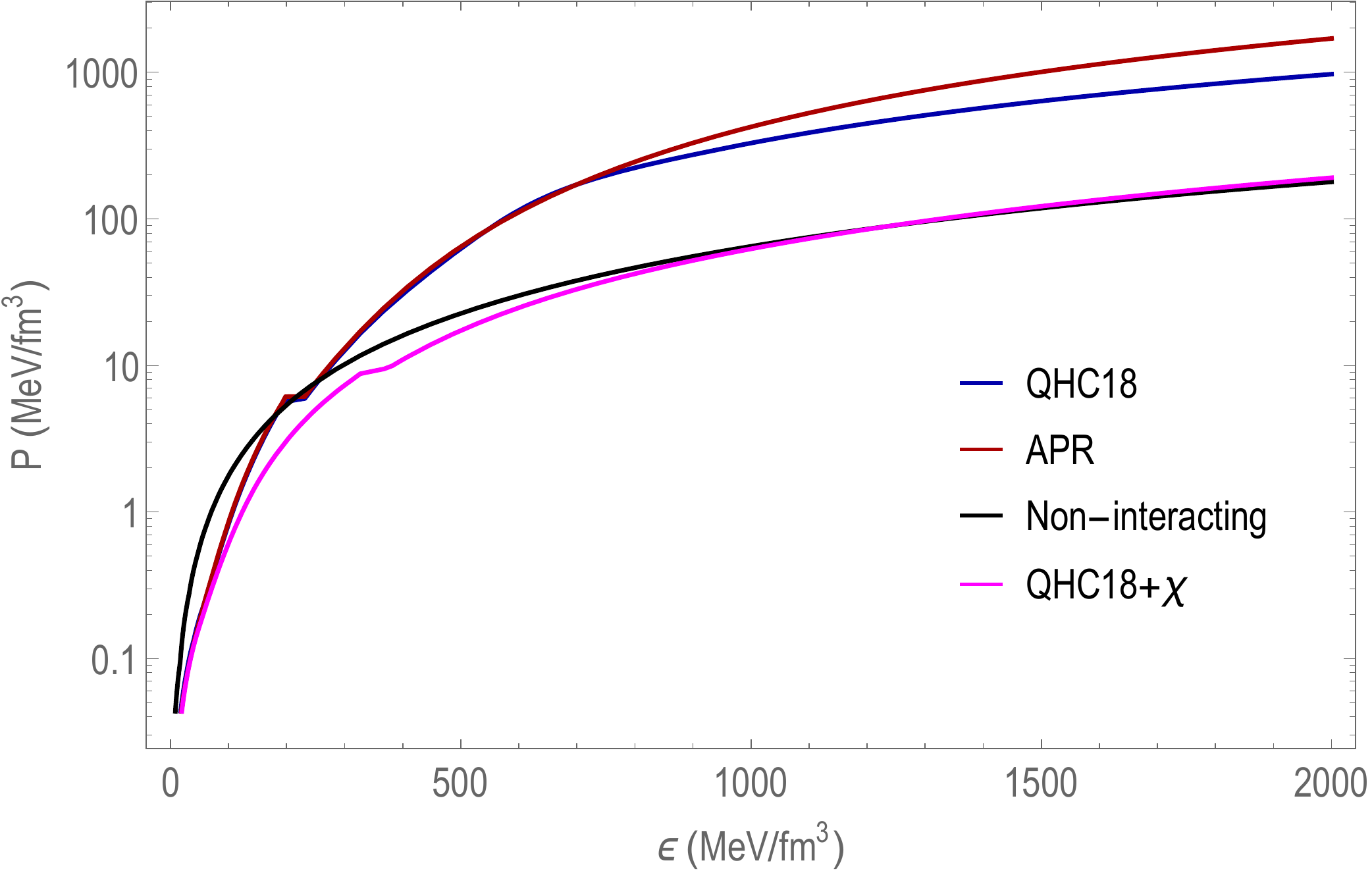}
\caption{\footnotesize{The equation of state for baryons coupled to $\chi$s, with $m_\chi=m_n$, and for 
the QHC18(0.8,1.5), APR and free neutron equations of state.   As explained in the text, at the resolution of this graph, the corresponding curves for baryons with the  QHC18 and APR equations of state in equilibrium with $\chi$s are indistinguishable.  Even though the APR equation of state is stiffer at higher densities than QHC18, in chemical equilibrium the core of the star would contain primarily dark fermions in either case.   At
nuclear matter density,  $\epsilon \simeq$
150 MeV/fm$^3$.}}
\label{fig-P}
\end{figure} 

   \begin{figure}[h]
\includegraphics[width = 0.45\textwidth]{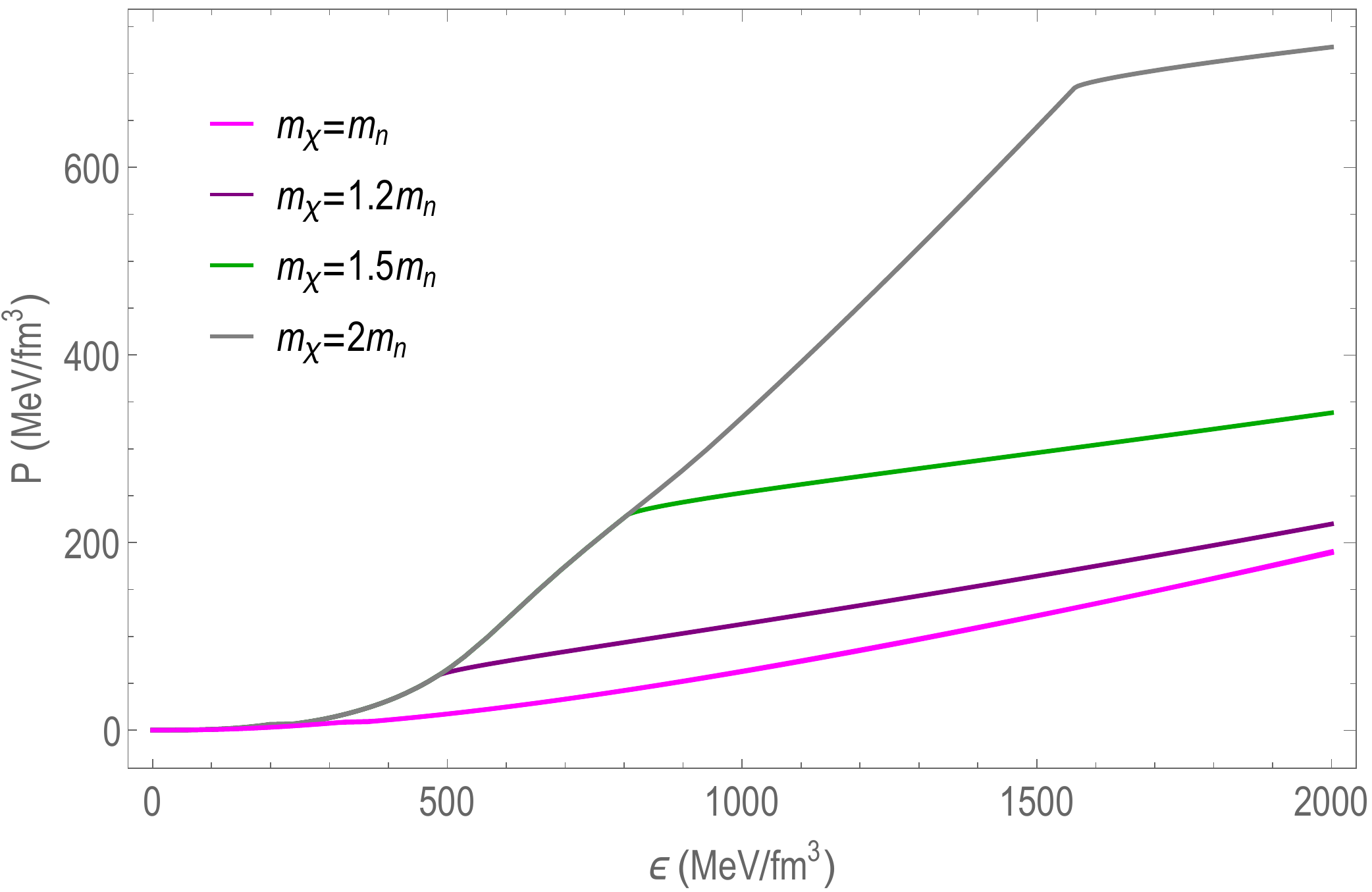}
\caption{\footnotesize{The QHC18(0.8,1.5) equation of state in chemical equilibrium with dark matter fermions of different masses: $m_\chi = m_n$, $m_\chi = 1.2m_n$, $m_\chi = 1.5m_n$, and $m_\chi = 2m_n$.  One sees here the dramatic softening induced by
    coupling to $\chi$s.  For larger $m_\chi$, the onset of the softening is pushed to higher $\epsilon$ where it becomes energetically favorable for a baryon to be converted to a $\chi$.}  }
\label{fig-Pdark}
\end{figure} 

    Figure~\ref{fig-P} shows the pressure, $P(\epsilon)$, calculated numerically for baryons and $\chi$'s in chemical equilibrium, for normal matter described by both the QHC18 and APR equations of state, as well as for these two equations of state without $\chi$, and for free neutrons.  The pressure of the equilibrium baryon-$\chi$ mixture does not depend noticeably on the baryon equation of state here.  
At high densities, where the pressure is lowered to essentially that of a free neutron gas, the matter is dominated by $\chi$s, while at densities below $2n_0$, the QHC18 and APR equations of state are identical by construction.   We show in Fig.~\ref{fig-Pdark}, the dependence of the 
total pressure of the coupled system for different $m_\chi$; the pressure follows the normal equation of state
up to energy densities where $\chi$'s are first allowed kinematically and then flattens.

\begin{figure}[h]
\includegraphics[width = 0.45\textwidth]{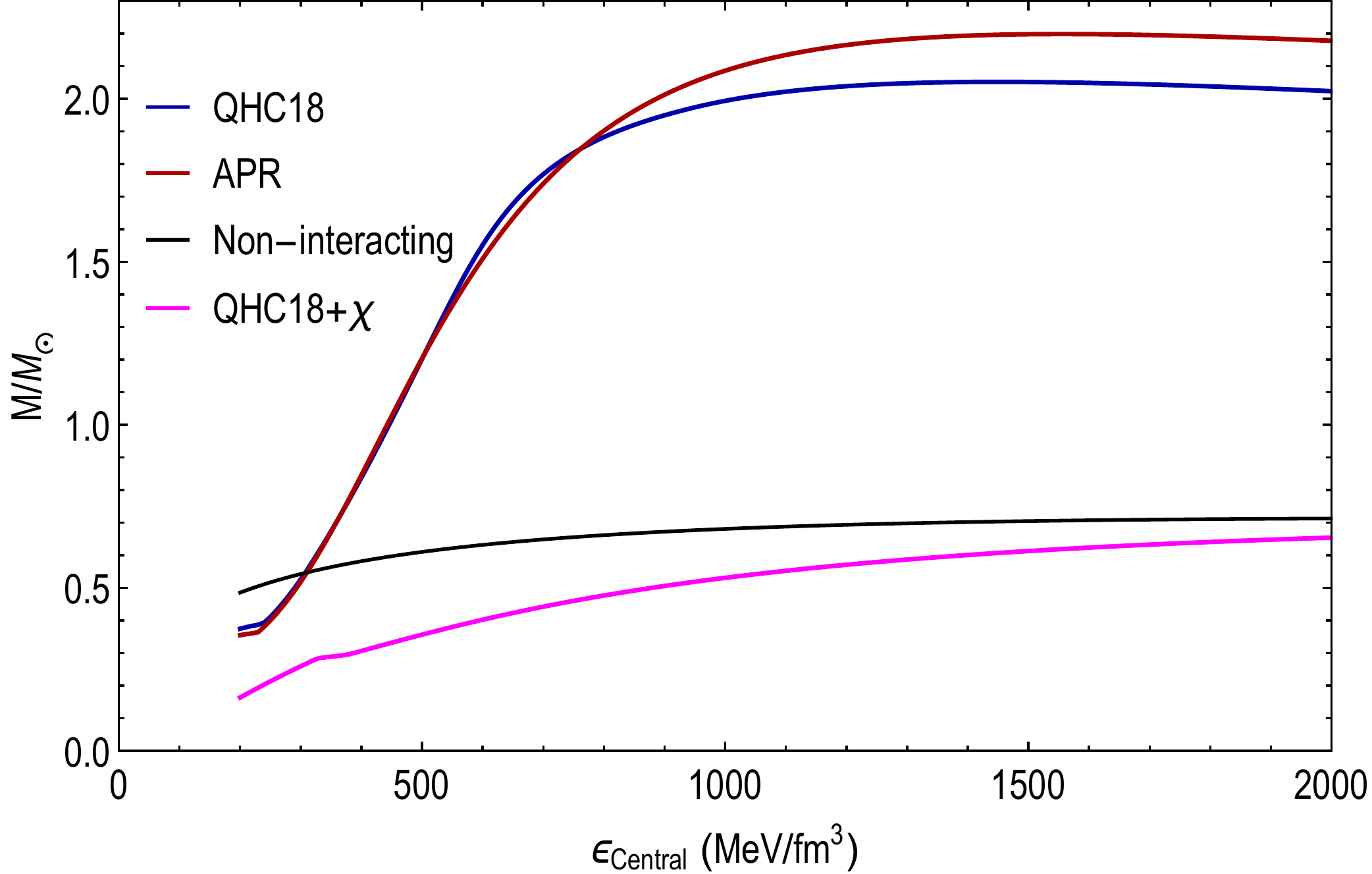}
\caption{\footnotesize{ 
 Neutron star masses vs. central density for baryons with the QHC18(0.8,1.5) equation of state in chemical equilibrium with $\chi$s having $m_\chi=m_n$,  and for the QHC18(0.8,1.5), APR, and free neutron equations of state.  This figure shows how
    coupling of baryons to weakly interacting dark matter precludes explanation of the existence of 
    neutron stars from 1-2$M_\odot$.  At the resolution of this graph, the corresponding curves for baryons with the QHC18 and APR equations of state in equilibrium with $\chi$s are indistinguishable.
}}
\label{mass}
\end{figure}

\begin{figure}[h]
\includegraphics[width = 0.45\textwidth]{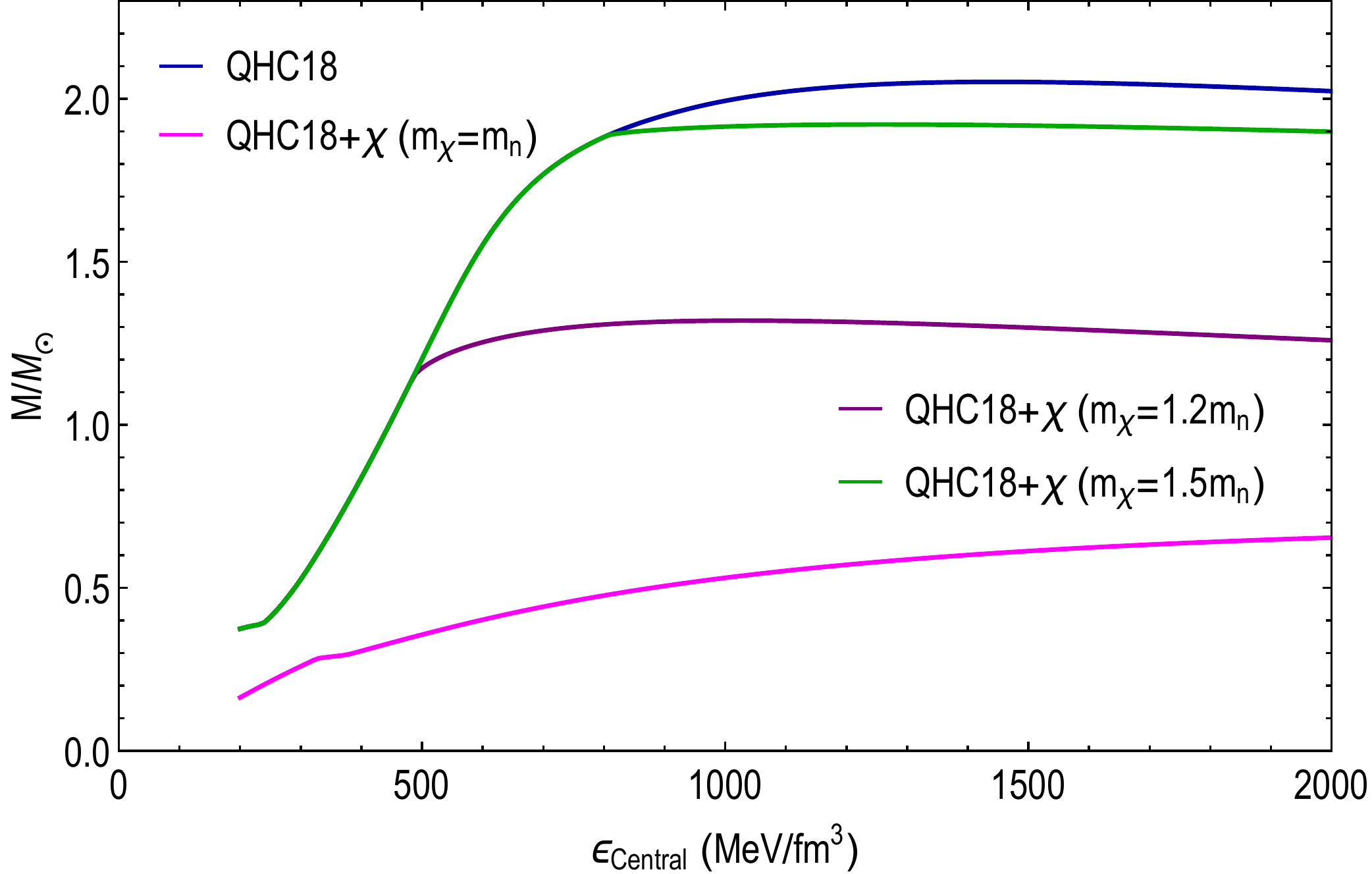}
\caption{\footnotesize{
Neutron star mass as a function of the central
energy density with baryons in chemical equilibrium with dark matter fermions for different masses: $m_\chi = m_n$, $m_\chi = 1.2m_n$, and $m_\chi = 1.5m_n$.   At larger $m_\chi$, conversion of baryons to dark matter is kinematically forbidden at lower densities, as shown in Fig. \ref{x}.  Therefore, the neutron star mass shown here is independent of the dark matter mass for sufficiently low central density.  For $m_\chi = 2m_n$, the neutron star mass is essentially unaffected by the small number of  $\chi$s present for this range of central densities.
}  }
\label{mass2}
\end{figure}

\begin{figure}[h]
\includegraphics[width = 0.45\textwidth]{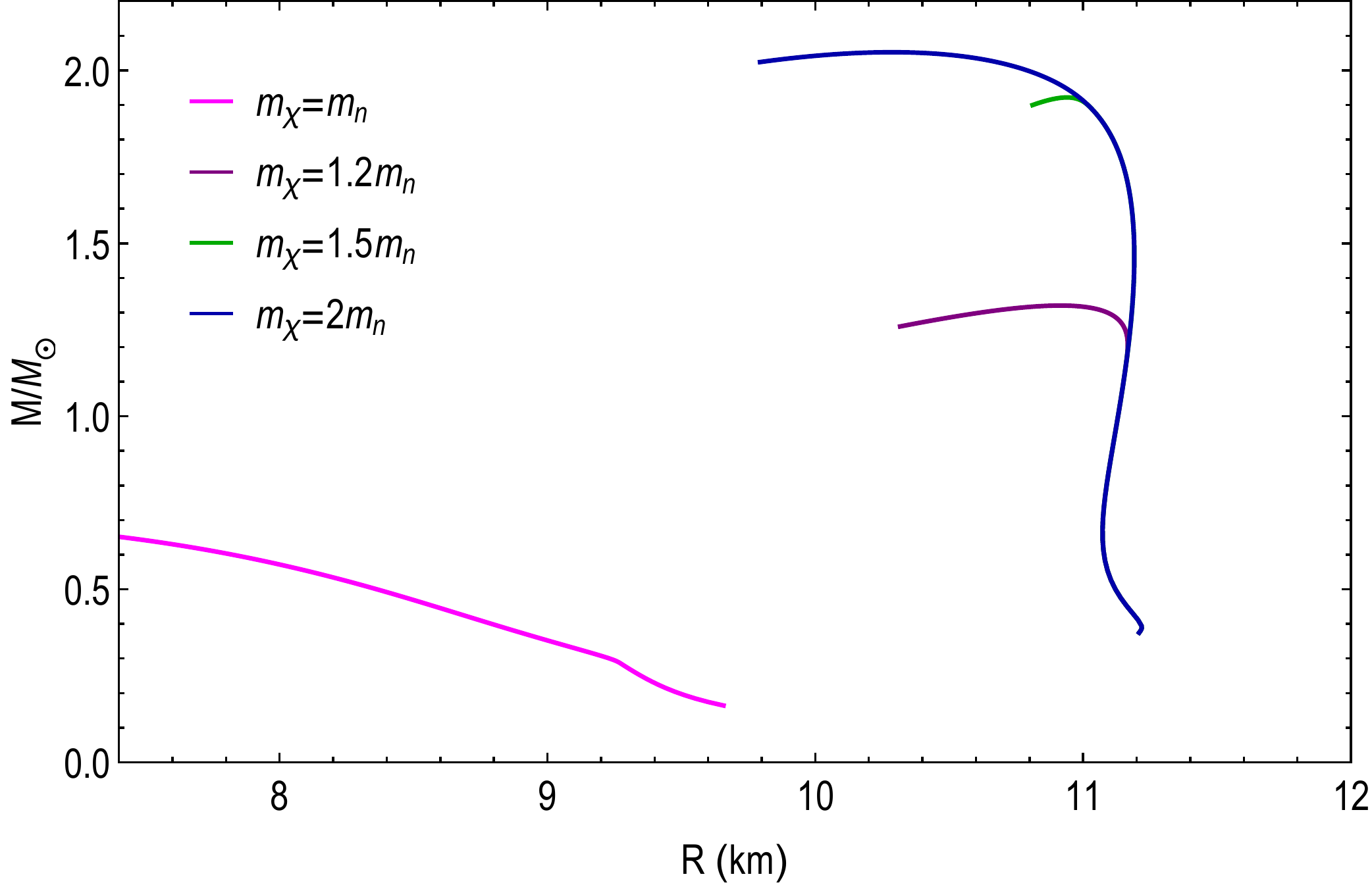}
\caption{\footnotesize{ Neutron star mass as a function of the neutron star radius for baryons with the QHC18 equation of state in chemical equilibrium with $\chi$s of mass $m_\chi = m_n$, $m_\chi = 1.2m_n$, and $m_\chi = 1.5m_n$.  Again, the QHC18 curves with no dark matter and for $m_\chi = 2m_n$ are essentially indistinguishable for this range of central energy densities.}   }
\label{mvsr}
\end{figure}

The resulting neutron star masses, found by
integrating the Tolman-Oppenheimer-Volkov equation \cite{tov1,tov2}, are shown in 
Fig.~\ref{mass} as a function
of the central energy density for QHC18 as well as for APR in equilibrium with $\chi$s for $m_\chi=m_n$ (again the results are indistinguishable at the resolution of the figure), for APR and QHC18 alone, and for free neutrons.
The maximum neutron star mass for the coupled matter is reduced from $\sim 2 M_\odot$ to $\sim 0.7 M_\odot$, even
below that for free neutrons; this reduction is a
consequence of the QHC18 equation of state at low densities being
softer than that of free neutrons (Fig.~\ref{fig-P}).  In Fig.~\ref{mass2} we show, 
for QHC18 coupled to $\chi$'s, neutron star masses for a range of $m_\chi$; in Fig.~\ref{mvsr}, 
we show the mass-radius relations for the same range of $m_\chi$.   As $m_\chi$ increases to $2m_n$,
the impact on the neutron star composition is negligible: there exist relatively few $\chi$'s and only at high
densities.   We conclude that the assumed coupling of baryons to non-interacting dark matter
lowers the maximum neutron star mass to well below that observed,
and thus the proposed exotic neutron decay mode is physically
untenable unless the dark matter equation of state satisfies very
demanding conditions, which we now discuss.

   In order to increase the mass of neutron stars with normal matter coupled to dark matter, the 
dark matter fermions would not only have to be strongly interacting among themselves, but the self-interactions would 
be strongly constrained.   Sizeable short-range dark matter
self-interactions are not impossible; the most stringent astrophysical
constraint comes from the Bullet Cluster, which allows cross-sections
$\sigma_{\chi} \lesssim (66/\mathrm{GeV})^2 \sim 10^{-24}$cm$^2$  -- on the scale of low energy baryonic cross sections -- for a particle with the
mass of a neutron \cite{Randall:2007ph}.   As a first scenario one might imagine that dark matter fermions exactly mirror normal matter.  Then the energy density and pressure in a neutron star in equilibrium with dark matter, with total fermion density, $2n_F$, would be just twice that of the normal matter alone at density $n_F$,
leading to a reduction in the maximum neutron star mass by $\sqrt2$ (see Appendix A of Ref.~\cite{nstar}), that is, a maximum mass of order 1.4$M_\odot$, below observations.  A second possibility could be that the dark matter has sufficiently repulsive interactions to overcome the softening of the equation of state due to adding a second species, thus accommodating 2$M_\odot$ neutron stars.   Such a scenario
makes very specific demands on the dark matter self-interaction strength as a function of the dark matter fermion density.   Construction of models that would yield the requisite self-interacting dark matter is left as a problem for the future.   

   If the neutron decays to multiple dark states, e.g., $n\to \chi +\phi $, where $\phi$ is a dark boson, our results hold when the $\phi$ do not carry a conserved charge and thus their chemical potential, $\mu_\phi$ must vanish;  if $\phi$ is a dark photon $\gamma_D$, for instance, obtaining $2M_\odot$ neutron stars would still require $\chi$s to have strong repulsive self-interactions.  Requiring the dark states to carry a conserved charge, and thus be part of a multi-particle dark sector, would allow $\mu_\phi \neq 0$ and possibly permit a resolution of the neutron lifetime puzzle as well as
the construction of dark matter equations of state consistent with  2$M_\odot$ neutron stars.   Such a scenario, an alternative to the dark matter sector having strong repulsive interactions, would also require some non-minimal multi-state dark sector.  

   Beyond the immediate motivation provided by the neutron lifetime puzzle, the present study demonstrates that neutron stars are powerful probes of baryon-violating $n$-$\chi$ couplings for $\chi$s as heavy as 2 GeV.      We stress that the present analysis considers only baryon number violating couplings, in contrast to analyses of dark matter capture by neutron stars which focus on elastic scattering of dark matter on baryons  \cite{Goldman:1989nd, Kouvaris:2007ay, Bertone:2007ae, deLavallaz:2010wp, Kouvaris:2010vv, McDermott:2011jp, Guver:2012ba, Bramante:2013hn, Bell:2013xk, Bertoni:2013bsa, Bramante:2013nma,Baryakhtar:2017dbj,  Raj:2017wrv}.  The class of interactions we consider here can easily have elastic baryon-$\chi$ cross-sections orders of magnitude below what can be tested with gravitational capture;  in the regime $m_\chi \lesssim m_n$ relevant for the neutron lifetime puzzle the bottle measurements stringently constrain the relevant coupling (implying $\sigma_{n\chi} \sim 10^{-54}$ cm$^2$).   Indeed, our results apply for all couplings large enough for baryon-$\chi$ conversion to reach equilibrium within the neutron star;  even conversion times of years, many orders of magnitude longer than that required to address the neutron lifetime puzzle, would lead to equilibrium.
   
    Non-zero strangeness in the quark matter phase in the interior of neutron stars would also allow analogous tests of a baryon-number-violating coupling of $\chi$ to hyperons.   As strange baryons are far less abundant in nature than neutrons, a $\chi$-hyperon coupling would be far more challenging to test in the lab, making neutron stars even more valuable probes.   

   As we have shown,  neutron stars can be used to constrain dark matter models in ways that are simply inaccessible
to other probes, whether cosmological or terrestrial, and thus provide a vital new window onto GeV-scale dark sectors.

\medskip

\noindent {\em Note added.} Since this paper was initially posted, several related works have appeared.  Notably, Refs.~\cite{sanjay, thomas} reach very similar conclusions to our own, while Ref.~\cite{epem} describes a search for neutron decays into  $\chi+e^+e^-$, with negative results; and Ref.~\cite{cline} builds an equation of state for self-interacting dark matter within a neutron star. \\ \\

\noindent {\bf Acknowledgments.} This research was supported in part by NSF Grants PHY-1506416 and PHY-1714042, and DOE Early Career Grant DE-SC0017840. 

\end{document}